\documentclass{epl}
\usepackage{amsfonts,amsmath,amssymb,graphicx}
\usepackage[english]{babel}

\newcommand{\ud}{\mathrm{d}}

\begin{document}
\title{The sediment of mixtures of charged colloids:\\ segregation and inhomogeneous electric fields}
\author{Jos Zwanikken \and Ren\'{e} van Roij}
\institute{Institute for Theoretical Physics, Utrecht University,
Leuvenlaan 4, 3584 CE Utrecht, The Netherlands}
\date{\textit{\today}}
\maketitle

\begin{abstract}
We theoretically study sedimentation-diffusion equilibrium of
dilute binary, ternary, and polydisperse mixtures of colloidal
particles with different buoyant masses and/or charges. We focus
on the low-salt regime, where the entropy of the screening ions
drives spontaneous charge separation and the formation of an
inhomogeneous macroscopic electric field. The resulting electric
force lifts the colloids against gravity, yielding highly
nonbarometric and even nonmonotonic colloidal density profiles.
The most profound effect is the phenomenon of segregation into
layers of colloids with equal mass-per-charge, including the
possibility that heavy colloidal species float onto lighter ones.
\end{abstract}

\section{Introduction}
Sedimentation in a suspension of colloidal particles is of
profound fundamental importance, and has been studied in detail
for a long time. For instance, in 1910 Perrin determined the
Boltzmann constant $k_B$ (and from this Avogadro's number) by
comparing the measured equilibrium density profile of a dilute
suspension with the theoretically predicted barometric law
\cite{perrin}, and more recently the full hard-sphere equation of
state was determined accurately from a single measurement of the
sedimentation profile of a dense suspension of colloidal hard
spheres \cite{piazza,rutgers}. In rather dilute suspensions of
charged colloids, however, strong deviations from the barometric
distribution have recently been theoretically predicted
\cite{biben1994,simonin,tellez2000,lowen,vanroij,biesheuvel},
experimentally observed \cite{albert,mircea1,paddy,mircea2}, and
simulated \cite{hynninen}, at least in the regime of extremely low
salinity. The most striking phenomenon is that the distribution of
colloids extends to much higher altitudes than is to be expected
on the basis of their buoyant mass. The force that lifts the
colloids against gravity is provided by an electric field that is
induced by spontaneous charge separation over macroscopic
distances. The corresponding cost of electrostatic and
gravitational energy is more than compensated by the gain of
translational entropy of the salt ions, which are distributed much
more homogeneously than when they would have had to follow the
colloids to the bottom part of the suspension. The same
entropy-induced lifting force was recently found to be responsible
for a so-called Brazil-nut effect in  binary mixtures of charged
colloidal particles under low-salt conditions: the heavy particles
can reside at higher altitudes than the lighter ones
\cite{esztermann}. In this Letter we extend the theoretical study
of sedimentation of charged colloids to binary, ternary, and
polydisperse mixtures, for which we calculate colloidal density
profiles and the entropy-induced electric field selfconsistently
within a Poisson-Boltzmann like density functional theory.

\section{Theory}
We consider an $n$-component suspension of negatively charged
colloidal spheres in a salt solution in the Earth's gravity field.
We label the colloidal species by $i=1,2,\cdots,n$, and denote the
electric charge, diameter, and buoyant mass for species $i$ by
$-Z_ie$, $\sigma_i$, and $m_i$, respectively, where $e$ is the
proton charge. We imagine the suspension to be in thermal and
osmotic contact with a reservoir that contains monovalent ions at
a concentration $2\rho_s$ and a solvent with dielectric constant
$\epsilon$ at temperature $T$. The ions are assumed to be massless
point particles with charge $\pm e$, and the solvent mass density
is taken into account by considering the buoyant instead of the
actual colloidal masses according to Archimedes. The volume of the
suspension is $V=AH$, with $A$ the (macroscopic) horizontal area
and $H$ the vertical height of the solvent meniscus above the
bottom of the system. The gravitational acceleration $g$  is in
the negative vertical direction. We are interested in the
equilibrium colloidal density profiles $\rho_i(x)$ as a function
of the altitude $x$ above the bottom at $x=0$ and below the
meniscus at $x=H$. The ion density profiles are denoted by
$\rho_{\pm}(x)$. In order to calculate these profiles we employ
the framework of density functional theory \cite{evans}, where the
equilibrium density profiles follow from the minimisation of a
grand potential functional $\Omega[\{\rho_i\},\rho_{\pm}]$ with
respect to all the profiles. Here we employ the mean-field free
energy functional that consists of entropic, gravitational, and
electrostatic contributions,
\begin{eqnarray}
\frac{\Omega}{k_BTA}&=&\sum_{\alpha=\pm} \int_{0}^H \ud x
\rho_{\alpha}(x)\big(\ln\frac{\rho_{\alpha}(x)}{\rho_s}-1\big)
+\sum_{i=1}^n \int_{0}^H
\ud x \rho_i(x)\big(\ln\frac{\rho_i(x)}{a_i}- 1\big)\nonumber\\
&+&\sum_{i=1}^n \int_{0}^H \ud x \frac{x}{L_i}\rho_i(x)
-\frac{2\pi\lambda_B}{2}\int_{0}^H \ud x\int_0^H\ud x'|x-x'|
Q(x)Q(x'), \label{fml:N}
\end{eqnarray}
where we introduced the fugacity (or activity) $a_i$ and the
gravitational length $L_i=k_BT/m_ig$ of species $i$, the Bjerrum
length $\lambda_B=e^2/\epsilon k_BT$, and the local charge density
$Q(x)=\rho_+(x)-\rho_-(x)-\sum_{i=1}^nZ_i\rho_i(x)$. The
electrostatic term follows from an in-plane integration of the
three-dimensional Coulomb law, $2\pi\int_0^{\infty}\ud R
R/\sqrt{|x-x'|^2 + R^2}=-2\pi|x-x'|$ with $R$ the in-plane
distance, where we ignored an irrelevant integration constant. The
functional (\ref{fml:N}) only couples the total charge density at
different heights in a mean-field fashion, i.e. electric double
layers and many other correlation effect are not taken into
account. In fact one easily checks that the thermodynamics of the
system reduces in the absence of gravity to an $(n+2)$-component
ideal-gas mixture, since then $Q(x)\equiv 0$ because of
translational invariance and charge neutrality. In the presence of
gravity, however, interesting structures already appear at this
relatively low level of sophistication. The Euler-Lagrange
equations $\delta\Omega/\delta\rho_{\alpha}(x)=0$ and
$\delta\Omega/\delta\rho_i(x)=0$ that must be satisfied by the the
equilibrium density profiles can be cast in the form
\begin{eqnarray}
\rho_{\pm}(x)&=&\rho_s\exp[\mp\phi(x)]; \label{fml:rhoa}\\
\rho_i(x)&=&a_i\exp[-x/L_i+Z_i\phi(x)];
\hspace{1cm}(i=1,2,\cdots,n)\label{fml:rhoi}
\end{eqnarray}
where $\phi(x)=-2\pi\lambda_B\int_0^H\ud x' Q(x')|x-x'|$ is the
dimensionless electrostatic potential gauged such that it is zero
in the reservoir where $Q(x)\equiv 0$. It turns out to be
convenient to rewrite it in differential form as the
Poisson-Boltzmann equation
\begin{equation}\label{fml:PB}
\frac{\ud^2 \phi(x)}{\ud x^2}\ =-4\pi\lambda_BQ(x)=\kappa^2 \sinh
\phi(x) + 4 \pi \lambda_B \sum_{i=1}^{n} Z_i \rho_i(x).
\end{equation}
Here we used Eq.(\ref{fml:rhoa}), and we defined the reservoir
screening constant $\kappa^2=8\pi\lambda_B\rho_s$. Together with
the boundary conditions $\phi'(0)=\phi'(H)=0$ (where a prime
denotes a derivative w.r.t. $x$) the Eqs.(\ref{fml:rhoi}) and
(\ref{fml:PB}) form a closed set of $n+1$ equations that can in
principle be solved to yield $\rho_i(x)$ and $\phi(x)$ for a given
thermodynamic state determined by the reservoir characteristics
$\kappa$ and $\lambda_B$, and by the colloidal fugacities $a_i$,
gravitational lengths $L_i$, and charges $Z_i$. In an experimental
situation, however, the total packing fraction
${\bar\eta}_i=(\pi/6)N_i\sigma_i^3/V$ of species $i$ is usually
fixed instead of the fugacity $a_i$, with $N_i$ the number of
colloids of species $i$. This conversion can easily be
accomplished here by regarding the fugacities $a_i$ as
normalization constants that take values such that
$\bar{\eta}_i=(1/H)\int_0^H\ud x \eta_i(x)$, with
$\eta_i(x)=(\pi/6)\rho_i(x)\sigma_i^3$ the local packing fraction
of species $i$. Note that the hard-core diameters $\sigma_i$ do
not appear at all in the Euler-Lagrange equations, since the
functional (\ref{fml:N}) ignores the hard-core part of the
(direct) correlations; we use for all colloidal species that
$\sigma_i=\sigma=150$nm only to be able to convert densities to
physically reasonable packing fractions. Once the profiles
$\phi(x)$ and $\rho_i(x)$ are known, then $\rho_{\pm}(x)$ follows
from Eq.(\ref{fml:rhoa}), and from this the total charge density
$Q(x)$. The magnitude of the electric field is given by
$E(x)=k_BT\phi'(x)/e$, where the prime denotes a derivative w.r.t.
the height $x$.

Unfortunately one cannot solve Eqs.(\ref{fml:rhoi}) and
(\ref{fml:PB}) analytically. However, its one-dimensional
character allows for a rather straightforward numerical solution
on a grid $\{x_k\}$ of heights, although some care must be taken
in dealing with the widely different length scales $H$ (say of the
order of centimeters) and $\kappa^{-1}$ (at most of the order of
microns) in realistic cases. In all our calculations we use a
non-equidistant grid that we adapt to the situation at hand,
making sure that it is fine enough to exclude numerical artifacts.
Typically we use 600 grid points for a system with a meniscus at
$H=20$cm. Our iterative scheme to solve Eqs.(\ref{fml:rhoi}) and
(\ref{fml:PB}) on the grid takes typically a few seconds on a
desktop PC for $n=1$, and a minute for $n=21$.

Before discussing our numerical results, we wish to point out that
the present theory reduces, for $n=1$, to the treatment of
Ref.\cite{vanroij}. For instance, the three regimes for the
colloidal density profile (barometric, linear, and exponential
with a large decay length) follow within the assumption of local
charge neutrality $Q(x)=0$, i.e.
$-\sinh\phi(x)=Z_1\rho_1(x)/2\rho_s\equiv y_1(x)$. In the linear
regime, where $1/Z_1<y_1(x)<1$, the potential is then from
Eq.(\ref{fml:PB}) linear in $x$, with a slope $\phi'(x)=1/Z_1L_1$
that is proportional to the electric field that lifts the colloids
against gravity \cite{vanroij}. This one-component result already
hints at an important complication for mixtures ($n\geq 2)$: since
generally $Z_iL_i\neq Z_jL_j$ a single electric field strength
cannot lift all the colloidal species simultaneously. On this
basis one could therefore already expect segregation in mixtures,
such that colloids with the same value $Z_iL_i$ (i.e. the same
charge-per-mass) are found at the same height. This is indeed what
our numerical results will show below.

\section{Segregation in binary and ternary mixtures}
We start our numerical investigation with a class of binary
mixtures ($n=2$) of equal-sized light ($i=1$) and heavy ($i=2$)
colloidal particles of various charge ratios.  We choose the
system parameters identical to those of the Monte Carlo
simulations of figure 3 of Ref.\cite{esztermann}: $\sigma=150$nm,
$H=1000\sigma$, $\lambda_B=\sigma/128$,
$L_1=\frac{3}{2}L_2=10\sigma$,
$\bar{\eta}_1=\bar{\eta}_2=(\pi/6)\times 10^{-4}$, $Z_1=15$, and
$Z_2$ varies between 15 and 45. The simulated system does not
contain added salt but only counterions, which we can represent
within our theory by the extremely low reservoir salt
concentration $\rho_s=1$nM, which we checked to be low enough to
be in the zero-added-salt limit as regards the colloidal profiles.
In Fig.1(a) we show, for $Z_2=45$, the density profiles of the two
colloidal species, as well as that of the counterions (and the
coions) in the inset. We observe profound colloidal segregation
into two layers, with the heavy species floating on top of the
lighter ones. The counterions are seen to be distributed
throughout the whole volume, i.e. much more homogeneously than
when all colloids would have been barometrically distributed in a
thin layer of thickness $~L_i$ just above the bottom (since in
that case the net ion charge would have been located in that same
thin layer). The resulting gain of ion entropy is the driving
force for the formation of the electric field that pushes the
heavy (highly charged) colloids to high altitudes against gravity
\cite{vanroij,hynninen}. In Fig.1(b) we show the mean height $h_i$
of species $i$, defined as
\begin{equation} \label{fml:hi}
h_i = \frac{ \int_{0}^H \ud x\ x\ \rho_i(x)}{\int_{0}^H \ud x\
\rho_i(x)},
\end{equation}
as a function of $Z_2$. We replot the Monte Carlo simulation
results of figure 3 of Ref.\cite{esztermann} (symbols), together
with the predictions for $h_i$ that follow from the present theory
(continuous curves). Given that there is not a single fit
parameter involved, the agreement is remarkable, certainly when
compared with the theoretical analysis on the basis exponentially
decaying density profiles as in Ref.\cite{esztermann}. Fig.1 shows
that the heavy particles are on top of the lighter ones,
$h_2>h_1$, provided $Z_2/Z_1\simeq 1.6$, a phenomenon that was
termed the "colloidal Brazil nut effect" in Ref.\cite{esztermann}.
For barometric profiles one would find that $h_i=L_i$, but we see
in all cases that $h_i\gg L_i$ due to the lift effect of the
induced electric field. The good agreement between theory and
simulation in Fig.1(b) also indicates that hard-core effects
(which are taken into account in the simulations but {\em not} in
the theory) are not so relevant, at least not in the parameter
regime studied here.
\begin{figure}[h]
\begin{center}
    \includegraphics[angle=270, scale= 0.3]{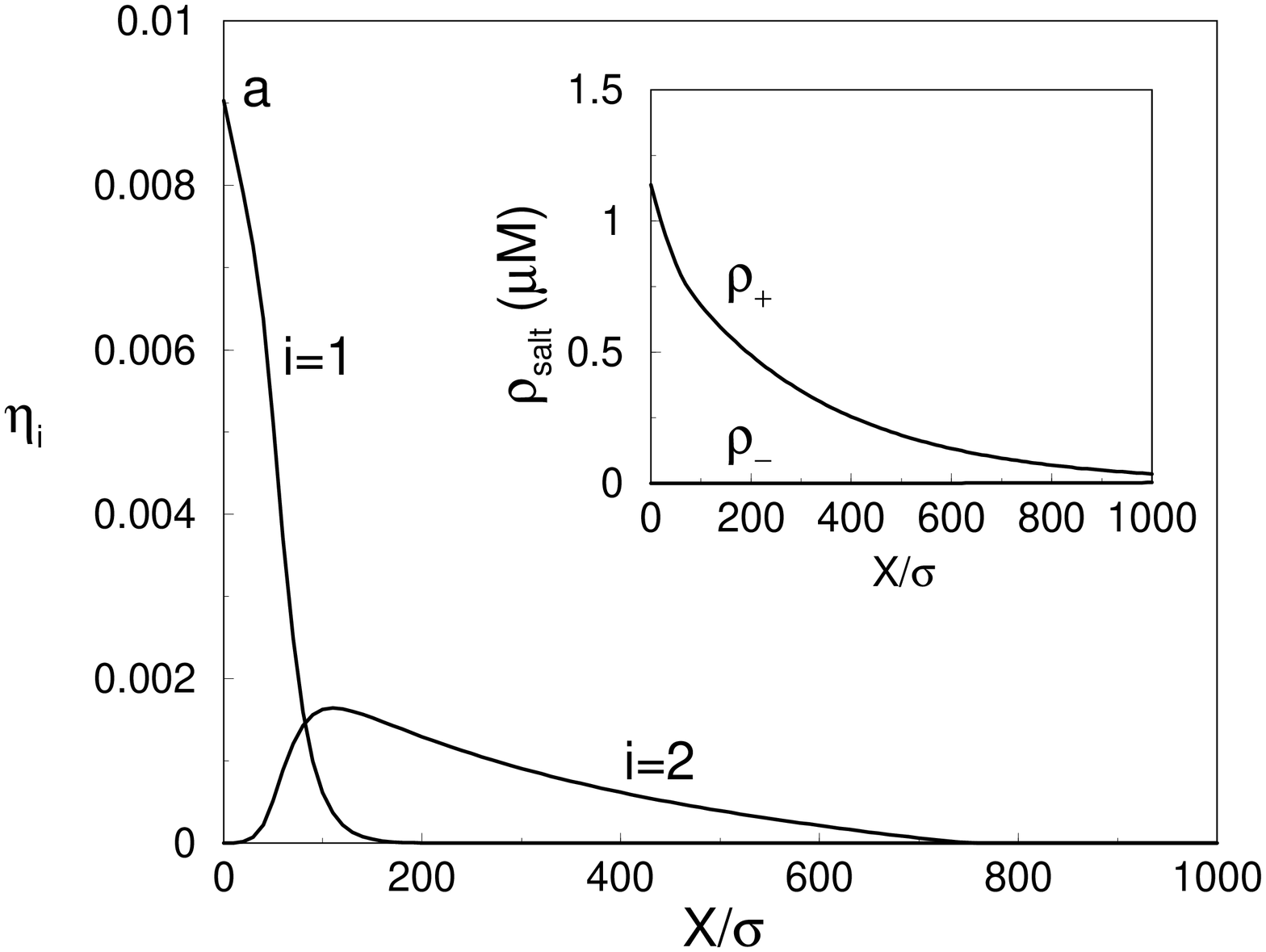}
    \includegraphics[angle=270, scale= 0.3]{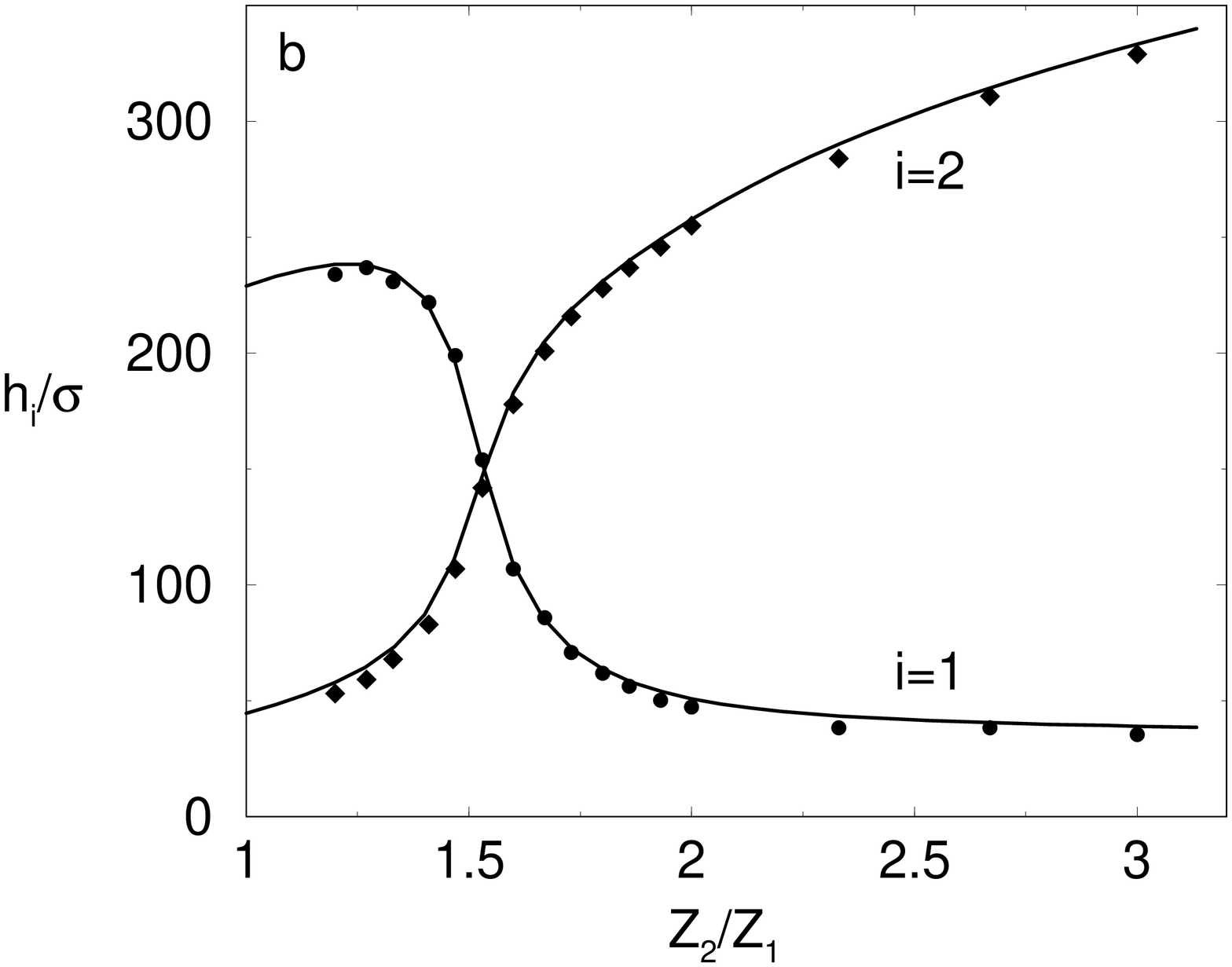}
     \caption{(a) Colloidal density profiles and counterion
     distribution (inset) for a deionised  binary mixture of
     equal-sized light ($i=1$) and heavy ($i=2$) colloidal
     spheres (diameter $\sigma=150$nm), with meniscus height
     $H=1000\sigma$, gravitational lengths $L_1=10\sigma$ and
     $L_2=20\sigma/3$, average packing fractions $\bar{\eta}_1=\bar{\eta}_2=(\pi/6)\times 10^{-4}$,
     colloidal charges $Z_1=15$ and $Z_2=45$, and Bjerrum length
     $\lambda_B=\sigma/128$. (b) Mean height $h_i$ of the two species for the same
     system as in (a) except now for a range of $Z_2$, showing a density inversion ($h_2>h_1$, the heavy particles
     floating on top of light ones) for
     $Z_2/Z_1>1.6$. The curves are predictions of the
     present theory, the symbols are simulation data of
     Ref.\cite{esztermann}.}
\label{fig:overgang}
\end{center}
\end{figure}

The next system we study is a ternary system ($n=3$) in a solvent
characterised by $\lambda_B=2.3$nm (ethanol at room temperature)
with a reservoir salt concentration  $\rho_s=10\mu$M and  meniscus
height $H=20$cm. The colloidal charges and gravitational lengths
are $Z_i=(1000,250,125)$ and $L_i=(1,2,1)$mm, respectively, and
the system is equimolar with  $\bar{\eta}_i=0.005$ for all three
species (recall that all species have the same diameter
$\sigma=150$nm). Fig. 2 shows the density profiles as predicted by
the present theory in (a), as well as the electric field in (b).
We find almost perfect segregation into three layers, such that
the mean heights satisfy $h_3<h_2<h_1$. Note that this ordering
coincides with the ordering $Z_3L_3<Z_2L_2<Z_1L_1$, and {\em not}
with the corresponding ordering of $L_i$ which one would expect on
the basis of a barometric distribution. In other words, this
system segregates according to mass-per-charge instead of the more
usual ordering according to mass: the colloids with the largest
mass-per-charge are found at the bottom. In the present case this
implies that the lightest colloids (species 2) are found in a
layer in between the equally heavy species 1 and 3. On the basis
of the one-component theory of Ref.\cite{vanroij} one would expect
a linearly decaying density profile of species $i$ in the layer
where species $i$ is dominant, as well as a linear electrostatic
potential. Denoting the derivative w.r.t. $x$ by a prime, the
one-component theory predicts
$\eta'_i(x)=\pi\rho_s\sigma^3/3Z_i^2L_i$ and $\phi'(x)=1/Z_iL_i$
in the layer with species $i$, corresponding to an electric field
strength $m_ig/Z_ie$ in this layer. These values for the density
gradients and the electric field are indicated in Fig.2(a) and
(b), respectively, and are in good agreement with the numerical
results. The result for the electric field can also be easily
obtained analytically for a mixture provided one assumes that
segregation takes place, such that $\rho_i(x)$ takes a maximum at
some height $x^*$ in the layer of species $i$. From
$\rho_i'(x^*)=0$ one obtains from Eq.(\ref{fml:rhoi}) that
$\phi'(x^*)=1/Z_iL_i$, in agreement with the numerical results.
The inset of Fig.2(b) shows the ratio of the total charge density
$Q(x)$ and the ion charge density $2\rho_s\sinh\phi(x)$, which is
such that $|Q(x)/2\rho_s\sinh\phi(x)|\ll 1$ for all $x$ except
close to $x=0$ and $x=H$ where it is $\sim1$. This indicates that
the system, with its inhomogeneous electric field, is yet
essentially locally charge neutral (but not exactly, and not at
all at the boundaries), suggesting that a description on the basis
of hydrostatic equilibrium within a local density approximation
should be rather accurate \cite{aldemar}.
\begin{figure}[h]
\begin{center}
\includegraphics[angle=270, scale= 0.3]{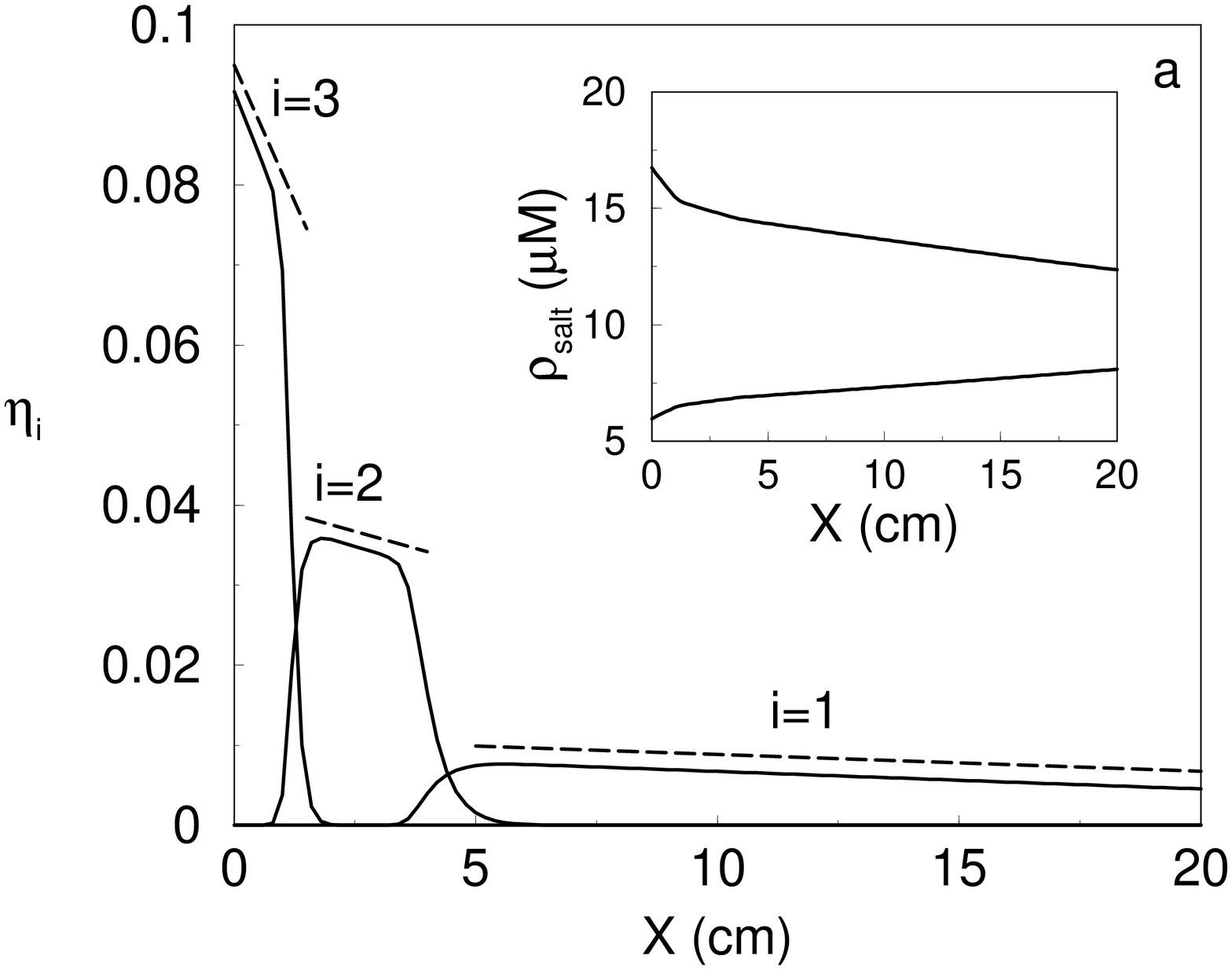}
\includegraphics[angle=270, scale= 0.28]{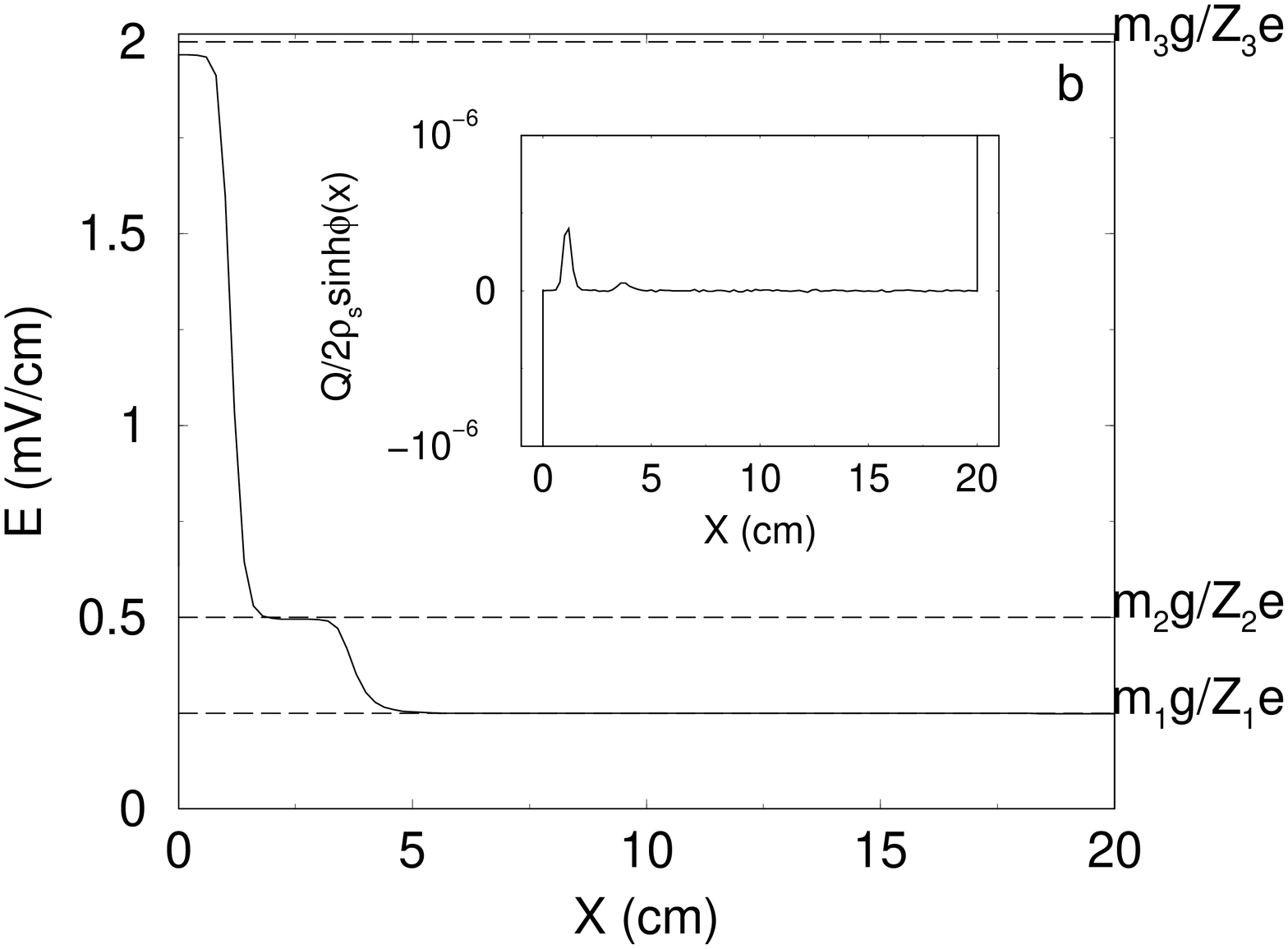}
\end{center}
     \caption{Density profiles (a) and the electric field profile
     (b) of an equimolar ternary mixture of equally-sized colloids
     with charges $Z_i=(1000,250,125)$ and gravitational lengths
     $L_i=(1,2,1)$mm with total packing fractions $\bar{\eta}_i=0.005$
     in a 20cm ethanol suspension at room
     temperature, with monovalent ionic strength $\rho_s=10\mu$M in the
     reservoir. Almost complete segregation takes place into
     well-defined layers of pure components, with
     a slope $\eta'_i(x)=\pi\rho_s\sigma^3/3Z_i^2L_i$
     and a constant electric
     field $m_ig/eZ_i$  in the layer with component $i$,
     as denoted by the dashed lines in (a) and (b), respectively.
     The inset in (b) shows a dimensionless measure for local charge neutrality (see main text).}
\end{figure}

\section{Polydisperse Mixtures}
We now extend our study to polydisperse mixtures, where one could
expect segregation into many layers on the basis of the results
for two and three components. We mimic the polydispersity by
considering a system of $n=21$ components, with $Z_i$ distributed
as a Gaussian with average of $250$ and a standard deviation of
50. This distribution is shown in the inset of Fig. 3(a), where
the vertical axis ($\bar{\eta}_i$) is proportional to the relative
frequency of species $i$ in the sample. We consider two
distributions for $L_i$: (A) $L_i=2$mm for all species, and (B)
$L_i=2\times (250/Z_i)^{3/2}$mm. Case B mimics the situation for
spheres of different size but the same mass density and surface
charge density, such that $Z_i$ is proportional to the surface
area and $L_i$ to the inverse volume of species $i$, i.e.
$L_i^2Z_i^3$ is a constant independent if $i$. The density
profiles, numerically obtained by solving the Eqs.(\ref{fml:rhoi})
and (\ref{fml:PB}) for $H=20$cm, $\lambda_B=2.3$nm,
$\rho_s=3\mu$M, and
$\bar{\eta}_{tot}=\sum_{i=1}^{21}\bar{\eta}_i=0.005$, are shown in
Fig.3 for all 21 components. Case A shows profound lifting and
layering, where the ordering is again determined by
mass-per-charge as illustrated by the three dashed curves (for
$Z_i=250$, 279, and 299) showing that the colloids in the
high-charge wing of the distribution reside at high altitudes.
Fig.3(b) shows the density profiles for case B, which does exhibit
lifting, but hardly any layering, and no density inversion at all.
This is completely consistent with the picture that the ordering
is determined by $Z_iL_i$, which for case B is such that
$Z_iL_i/Z_jL_j=\sqrt{Z_j/Z_i}$, i.e. the highly-charged particle
are expected at the bottom while the relative spread in $Z_iL_i$
is relatively small compared to case A, where
$Z_iL_i/Z_jL_j=Z_i/Z_j$. The inset of (b) shows the total packing
fraction profiles $\eta_{tot}(x)=\sum_{i=1}^{21}\eta_i(x)$ of both
case A and B together with the one-component profile ($n=1$) with
$Z_1=250$ and $L_1=2$mm at $\bar{\eta}_1=0.005$. Perhaps
surprisingly there is hardly any distinction between the pure
system and case B, whereas there is a small difference with case
A. These profiles show that the main distinction between these
polydisperse systems and the underlying one-component one concerns
the layering phenomenon (provided $Z_iL_i$ varies sufficiently for
all the species), and {\em not} the total distribution of the
colloids. Perhaps this fractionation effect could be exploited
experimentally to purify a polydisperse mixture.

%%%% PLAATJE %%%%

\begin{figure}[h]
\begin{center}
\includegraphics[angle=270, scale= 0.3]{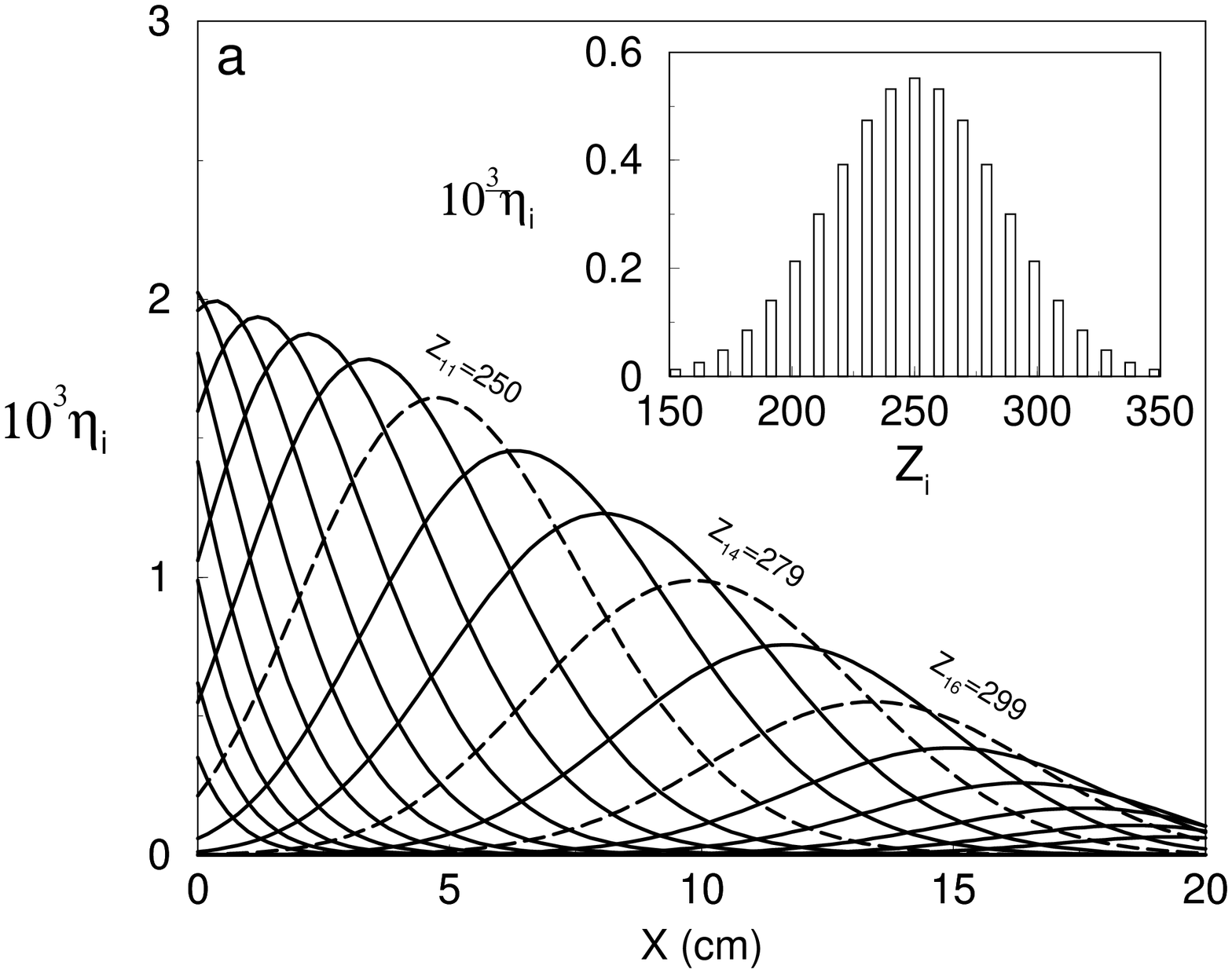}
\includegraphics[angle=270, scale= 0.3]{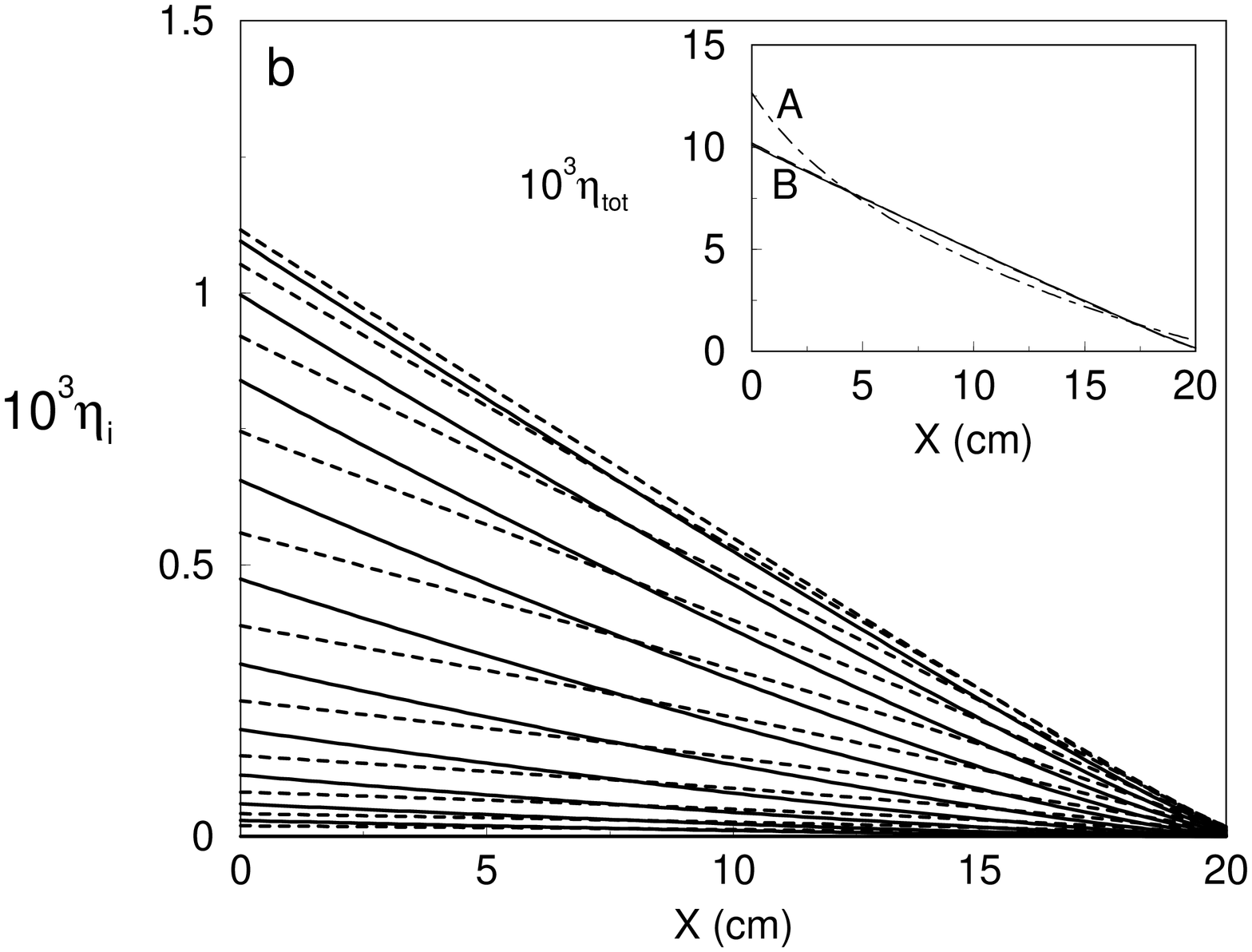}
\end{center}
     \caption{Density profiles of a 21-component colloidal suspension
       with a Gaussian charge distribution as
     illustrated in the inset of (a), at total packing
     fraction $0.005$, reservoir salt concentration $\rho_s=3\mu$M,
     and Bjerrum length $\lambda_B=2.3$nm. The gravitational
     lengths are as in case A (see main text) in (a), and case B in (b).
     In (b) the solid curves are for $Z_i>250$, and the dashed ones for $Z_i\leq 250$.
     The inset of (b) shows the total
     packing fraction profile for case A and
     B, together with that of the underlying one-component
     system (see main text).
} \label{fig:CDE}
\end{figure}

\section{Conclusions and discussion}
We have studied sedi\-men\-tation equilibrium of $n$-compo\-nent
systems of charged colloidal  particles at low salinity within a
Poisson-Boltzmann-like density functional. The Euler-Lagrange
equations that describe the minimisation of the functional are
solved numerically on a one dimensional grid of heights for a
number of system parameters. For $n=1$ the theory reduces to the
one-component studies as presented in Ref.\cite{vanroij,hynninen},
and for $n=2$ we quantitatively reproduce the simulation results
of Ref.\cite{esztermann}, where density inversion was found such
that the heavier colloids float on top of the lighter ones. This
effect is caused by a self-consistent electric field that lifts
the higher charged (heavy) particles to higher altitudes than the
lower charged (lighter) colloids. We show that the layering of the
colloids according to mass-per-charge can persist for ternary
($n=3$) as well as for polydisperse (here $n=21$) mixtures. Given
the good account that the present theory gives for simulations
\cite{hynninen} and experiments \cite{mircea1,paddy,mircea2} of
one-component systems, and for the simulations of binary systems
\cite{esztermann}, it is tempting to argue that the theory is also
(qualitatively) reliable for ternary or polydisperse systems, i.e.
the predicted segregation and layering should be experimentally
observable. One should bare in mind, however, that the present
theory ignores the hard-core of the colloidal particles, and is
therefore expected to break down at higher packing fractions. It
also ignores effects due to charge renormalisation, which becomes
relevant when $Z_i\lambda_B/\sigma_i\gg 1$. Work on extending the
theory in these directions is in progress
\cite{biesheuvel2,aldemar}.

All results presented in this paper were obtained with the
zero-field boundary conditions $\phi'(0)=\phi'(H)=0$. We checked
explicitly, however, that other boundary conditions that respect
global charge neutrality, such as
$\phi'(0)=\phi'(H)=eE_{ext}/k_BT$ (describing a suspension in a
homogeneous external electric field $E_{ext}$) or
$\phi(0)=\phi(H)$ and $\phi'(0)=\phi'(H)$ (describing a
short-circuited bottom and meniscus) give indistinguishable
density profiles, except in two layers of thickness
$\sim\kappa^{-1}\ll 10\mu$m in the vicinity of the bottom and the
meniscus. This insensitivity to the boundary conditions is not
surprising in the light of the fact that the whole phenomenology
in these systems is driven by the entropy of the microscopic ions,
i.e. a bulk contribution to the grand potential that should
dominate any boundary (surface) contribution.

\section{Acknowledgements}
It is a pleasure to thank Maarten Biesheuvel for interesting discussions.
This work is part of the research programme of the "Stichting voor
Fundamenteel Onderzoek der Materie (FOM)", which is financially
supported by the "Nederlandse organisatie voor Wetenschappelijk Onderzoek
(NWO)".

\end{document}